# Experiments on route choice set generation using a large GPS trajectory set

Rui Yao, Shlomo Bekhor

Department of Civil and Environmental Engineering,
Technion – Israel Institute of Technology, Haifa 32000, Israel

## ABSTRACT

Several route choice models developed in the literature were based on a relatively small number of observations. With the extensive use of tracking devices in recent surveys, there is a possibility to obtain insights with respect to the traveler's choice behavior.

In this paper, different path generation algorithms are evaluated using a large GPS trajectory dataset. The dataset contains 6,000 observations from Tel Aviv metropolitan area. An initial analysis is performed by generating a single route based on the shortest path. Almost 60% percent of the 6,000 observations can be covered (assuming a threshold of 80% overlap) using a single path. This result significantly contrasts previous literature findings.

Link penalty, link elimination, simulation and via-node methods are applied to generate route sets, and the consistency of the algorithms are compared. A modified link penalty method, which accounts for preference of using higher hierarchical roads, provides a route set with 97% coverage (80% overlap threshold). The via-node method produces route set with satisfying coverage, and generates routes that are more heterogeneous (in terms number of links and routes ratio).

# INTRODUCTION

Route choice modeling provides the probability that an individual traveler chooses a certain route from a set of available alternative routes. Most existing studies on route choice typically consider a path-based two-step approach (Ben-Akiva and Bierlaire, 2003; Bekhor et al., 2006). First, possible alternative paths are explicitly generated to form the choice set. Then, the pre-defined choice set is used for model estimation. Some other studies proposed implicit link-based approaches to model the route choice behavior as a recursive link choice behavior, without explicitly specifying the choice set (Dial, 1971; Fosgerau et al., 2013).

By explicitly specifying the set of alternative routes, we can examine possible selection criteria for generating realistic routes. However, route choice set generation is still challenging to find a representative set not only for model estimation, but also for applications in travel behavior analysis.

Several methods based on variations on the shortest path were developed to generate route choice sets. Typically, shortest paths are iteratively generated after variating link impedances. Ben-Akiva et al. (1984) proposed the labelling approach, in which different link attributes are exploited to formulate different generalized link costs, and used as selection criteria for alternative routes. Dial (2000) generalized the labeling approach and proposed to construct efficient paths, which minimize a linear combination of label costs.

De la Barra et al. (1993) proposed the link penalty method, which gradually increases the impedance of all links on the shortest path, and repeat the process till no new path can be found. Link elimination was proposed by Azevedo et al. (1993), in which links on the shortest path are removed. The main concern regarding link elimination method is network disconnection, Prato and Bekhor (2006) suggests a variant to remove only one link at each iteration, if the link takes the driver further away from destination or to lower hierarchical roads. Simulation method generates alternative paths by drawing link impedances from different probability distributions in each iteration, the algorithm then runs pre-defined number of draws to produce alternative paths.

Other studies incorporate network topology in route set generation. For example, Abraham et al. (2013) introduced the via-node method to find alternative routes for navigation, by finding shortest paths that pass through some via-nodes. Candidate via-nodes are meeting nodes in the bidirectional search which start from the origin and from destination respectively. Via-nodes are selected based on the idea of "admissible path", where these paths are sufficiently different from the shortest path, with no unnecessary detours, and not locally over circuitous. Luxen and Schieferdecker (2014) extended the via-node method by selecting natural graph cuts as candidate via-node set, and improved the algorithm performance.

Table 1 provides an overview of the performances of different algorithms (car mode) in terms of coverage measurement, which is an indication of the algorithm ability to reproduce the observed routes (discussed in detail later). In general, with higher resolution network, map matching of GPS trajectories on road network becomes substantially more time consuming.

In contrast to most studies, which use either a detail network or a relatively larger observation set, we use a detail network and a large GPS trajectory set.

Table 1 Performances of path generation algorithms (car networks)

| Method | Literature | Network Size | # of observations | Number of Routes | Coverage (80% overlap threshold, or best coverage obtained) |
|---|---|---|---|---|---|
| Labeling Approach | Bekhor et al. (2006) | about 13,000 nodes, about 34,000 links | 188 | 1 | 46% |
| | Prato and Bekhor (2007) | 419 nodes, 1,427 links | 236 | 1 | 31% |
| | Spissu, et al. (2011) | 18,000 links | 393 | 1 | 47% |
| | Quattrone and Vitetta (2011) | 4480 nodes, 16,029 links | 332 | 5 | 75% |
| | Zhu and Levinson (2015) | 8,618 nodes, 22,477 links | 657 | 1 | 23% |
| | Tang and Levinson (2018) | 8,618 nodes, 22,477 links | 124 | 1 | 28% |
| Link Elimination | Bekhor et al. (2006) | about 13,000 nodes, about 34,000 links | 188 | 30 | 71% |
| | Frejinger and Bierlaire (2007) | 3077 nodes, 7459 links | 2,978 | 15 | 80% |
| | Prato and Bekhor (2007) | 419 nodes, 1,427 links | 236 | 10 | 70% |
| | Pillat et al. (2011) | 7,703 nodes, 22,620 links | 1,089 | 1-13 | 60% |
| | Ding et al. (2014) | 7,808 nodes, 11,106 links | 997 | 30 | 79% |
| | Rieser-Schüssler, et al. (2013) | 408,636 nodes, 882,120 links | 500 | 100 | 75% |
| Link Penalty | Bekhor et al. (2006) | about 13,000 nodes, about 34,000 links | 188 | 40 | 80% |
| | Prato and Bekhor (2007) | 419 nodes, 1,427 links | 236 | 15 | 62% |

In this study, we perform experiments on route choice set generation, applying classical methods and the recent via-node method. Specifically, link penalty, link elimination, simulation and via-node methods are applied to generate route sets, and the consistency of the algorithms are compared using a large GPS trajectory set.

## METHODOLOGY

This section describes the methodology, consisting of: (i) survey and dataset description, (ii) application of the path generation algorithms, (iii) consistency measurement

**The Tel Aviv Metropolitan Survey Data**

This paper uses a dataset from the household survey conducted in the Tel Aviv metropolitan area between September 2016 and December 2017. The survey collected general information about the household members and their activities. GPS data loggers were provided to the household members, which recorded their locations for 48 consecutive hours with 2 second time step on average.

The overall sample includes 28,530 individuals living in 10,305 households. A total of 233,588 trips were recorded, in which 59.3% of them is car (either driver or passenger) trips, 25.9% is walking trips, 10.9% is public transit trips, 2.5% bicycle trips and 1.4% is motorcycle trips.

Out of all the trips, we are interested in car trips with sufficient trip length and are related to main activities, which are commute trips, maintenance trips (e.g. shopping) and personal trips

(visiting friends, entertainments, etc.). We set a minimum of 2 km as threshold, which results in 38,175 main activity car trips.

There are 7.3 billion raw GPS readings corresponding to all trips. After performing logical checks and deleting observations with gross errors in GPS for the selected car trips, there are 3,363,755 points related to 6,000 car trips performed by 2,739 persons. The logical checks include that (a) a trip should originate and end in Tel Aviv Metropolitan area, and (b) the average distance between two consecutive GPS data point is no more than 100 m, which corresponds to the minimum link length in the Tel Aviv network.

After cleaning and filtering the GPS data, the 6,000 car trips are map matched to a detailed planning network of Tel Aviv metropolitan area, which contains 8,583 nodes and 21,151 directed links. We apply Hidden Markov Model (Newson and Krumm, 2009) to perform map matching, by assuming the GPS data noise follows Gaussian Distribution with mean $\mu = 0$ and standard deviation $\sigma = 20$, tolerance for non-direct route $\beta = 2$ and the maximal search distance for candidate road segment is 50 m. In addition, manual route inspection on the matched routes was performed to ensure the route quality.

**Path Set Generation**

We apply variants of selective path generation algorithms based on shortest path to generate alternative paths, they are: Labeling Approach, Link Penalty, Link Elimination, Simulation Method, in which these algorithms repeatedly change link impendence and find the minimum cost path. Moreover, we apply the Via-Node Method, which finds admissible paths that pass through via nodes.

*Labeling Approach*

Different labels which are related to the time-dependent link travel time are used to generate alternative routes: fastest routes using link travel times correspond to AM-peak, PM-peak, off-peak, free-flow and trip-specific departure period; and shortest distance routes. In this method, a single route is generated for each label.

*Link Penalty*

Two variants of link penalty methods are applied by both replicating for 25 and 50 iterations, with the following procedures: (a) calculation of the fastest routes using trip-specific departure period link travel times; (b) increase travel times on the shortest path links by a factor; (c) calculation of the fastest routes and comparison with the existing routes in the path set.

The first variant applies a uniform factor of 1.05 to all the links on the shortest path. While the second variant also applies penalties to all the links on the shortest path, but with a higher penalty factor of 1.20 on local streets and a lower factor of 1.05 on higher hierarchical roads. Because we found that most of the observations prefer using the higher hierarchical roads (e.g. highways and urban arterials) even if the total travel time is higher.

*Link Elimination*

We apply link elimination as proposed by Prato and Bekhor (2006) by repeating for 25 and 50 iterations the following three steps: (a) computation of the fastest path by considering the trip-specific departure period travel time, (b) elimination of a link belonging to the current shortest path, and (c) computation of the next shortest path. Shortest-path links are eliminated if they take the driver farther from the destination or compel the driver to turn from a high hierarchical road to a low hierarchical road.

*Simulation Method*

Two simulation approaches are implemented by computing the shortest path for each draw of link impedances of the trip-specific departure period. The two approaches exploit the same procedure to draw impedances from a truncated normal distribution characterized from the following parameters: (a) mean equal to the travel time, (b) variance equal to a percentage of the mean, (c) left truncation limit equal to the free-flow time, and (d) right truncation limit equal to the travel time calculated for a minimum speed assumed equal to 10 km/h.

The first simulation approach sets the variance equal to 20% of the mean. The second simulation approach defines the variance equal to the mean. Both simulation approaches are extracted for 25 and 50 draws.

*Via-Node Method*

We follow the procedures proposed by Abraham et al. (2013) to generate alternative routes. The procedure starts by a bidirectional Dijkstra search to grow the search spaces from origin (s) and destination (t) respectively, with maximum search distance $(1+\epsilon)d_{st}$ where $\epsilon$ is the maximum detour percentage and $d_{st}$ is the s-t shortest path distance. And the meeting nodes in search spaces are candidate via nodes (Figure 1).

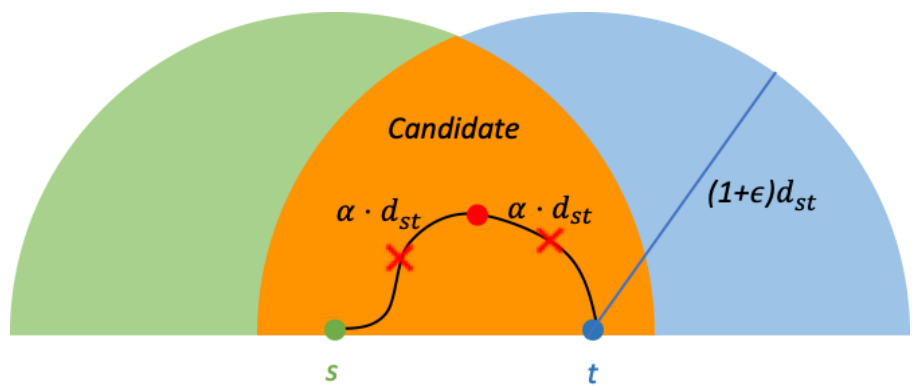

Figure 1 Candidate Via Nodes

In addition to the maximum detour constraint, the algorithm aims to generate routes with overlap length to the shortest path no more than $\gamma \cdot d_{st}$, i.e. the route generated is sufficiently different from the shortest path. Moreover, the route also should be reasonable, and it is formally defined as $\alpha$-locally optimal. That is, the sub-path with forward and backward length $\alpha \cdot d_{st}$ from the via-node should be a shortest path. For example, in Figure 1 the sub-path between the red crosses should be a shortest path.

The maximum detour factor $\epsilon$ is set to be 30%, the overlap factor $\gamma$ to be 85% and local optimality factor $\alpha$ to be 20%, these parameters are the best setting obtained by trial and error with our dataset.

**Consistency Measurement**

The effectiveness of route choice set generation method is evaluated by the generated routes' coverage over the observed routes. The coverage is the percentage of observations for which the algorithm is able to generate a route that satisfies a threshold for the overlap measure (eq. 1):

$$\max_{r} \sum_{n=1}^{N} I(O_n \geq \delta) \quad (1)$$

where $I(\cdot)$ is the coverage function, and it equals to 1 if the argument is true, $O_n$ is the overlap percentage of observation $n$ and $\delta$ is the threshold for the overlap measure. The overlap measure evaluates the consistency of a generated route with respect to the observed behavior by considering the length of the links shared between generated and observed routes (eq. 2):

$$O_n = \frac{L_{nr}}{L_n} \quad (2)$$

in which, $L_{nr}$ is the overlapping length of generated route $r$ for observation $n$, and $L_n$ is the length of the observed route for observation $n$.

# RESULTS

Several variations of the route generation algorithms described above were examined in this study. Table 2 shows the coverage results of different shortest path labels. That is, each algorithm generates exactly one route by minimizing a specific time-period travel time or distance.

Table 2 Percentage Coverage of the Labeling Approach

| Label | Percentage Coverage | | | |
| --- | --- | --- | --- | --- |
| | 80% Overlap Threshold | 90% Overlap Threshold | 100% Overlap Threshold | Average Overlap |
| Fastest Path (Trip-specific departure period) | 59% | 48% | 37% | 75% |
| Fastest Path (AM peak) | 57% | 46% | 35% | 74% |
| Fastest Path (PM peak) | 59% | 48% | 37% | 75% |
| Fastest Path (Off peak) | 59% | 49% | 38% | 75% |
| Fastest Path (Free-flow) | 57% | 47% | 37% | 73% |
| Shortest Distance Path | 42% | 35% | 32% | 61% |

The above table shows that either the trip-specific departure period fastest path or the off-peak fastest path are the best labels. Almost 60% percent of the 6,000 observations can be covered (assuming a threshold of 80% distance overlap). This result is significantly higher in comparison to previous results reported in the literature, and may be related to compliance with software navigation apps (such as Waze and Google Map).

As indicated above, the labeling approach is able to cover almost 60% of the observations. In order to improve the coverage, additional routes must be generated. We examine link penalty, link elimination, simulation and via-node methods to generate multiple paths. The results are presented in Table 3.

Table 2 Coverage of selected path generation algorithms

| Path Generation Algorithm | 80% Overlap Threshold | 100% Overlap Threshold | Average Overlap | Avg. number of unique routes |
|---|---|---|---|---|
| **25 Shortest Path Iterations** | | | | |
| Link Penalty (uniform factor) | 94% | 66% | 96% | 11.95 |
| Link Penalty (higher penalty on local roads) | 97% | 67% | 97% | 14.53 |
| Link Elimination | 82% | 56% | 91% | 15.79 |
| Simulation (large variance) | 74% | 53% | 85% | 2.94 |
| Simulation (small variance) | 74% | 53% | 85% | 2.95 |
| Combined | 99% | 76% | 98% | 32.34 |
| **50 Shortest Path Iterations** | | | | |
| Link Penalty (uniform factor) | 96% | 69% | 97% | 26.64 |
| Link Penalty (higher penalty on local roads) | 98% | 69% | 97% | 31.68 |
| Link Elimination | 83% | 56% | 91% | 23.58 |
| Simulation (large variance) | 77% | 55% | 86% | 3.5 |
| Simulation (small variance) | 77% | 56% | 86% | 3.5 |
| Combined | 99% | 79% | 98% | 66.55 |
| **Via-node Method** | | | | |
| Via-Node | 94% | 69% | 96% | 28.26 |

The link penalty method produces a very good coverage (94% for 80% threshold). The coverage improves to 97% if local roads are more heavily penalized in comparison to higher hierarchy roads. Also, the via-node method provides satisfying results in terms of coverage, with similar average number of unique routes compared to the 50 iteration link penalty methods. Combining all 5 methods, with 25 iterations for each method, we obtain an excellent coverage (99% for 80% threshold). Note that on average, for a total of 125 iterations, the combined route set contains 33 routes on average.

To visualize the consistency of the applied algorithms regarding the observed behavior, Figure 2 shows the distribution of coverage over the cumulative percentage of observations with respect to different path generation algorithms with 25 iterations (except via-node method). The area under each line measures the consistency of each algorithm (Prato and Bekhor, 2006), and it is equal to the average overlap. The consistency of the applied methods varies from 75% for the labeling approach to 97% for the link penalty method.

There is a trade-off between the number of iterations (or number of routes generated) in terms of computational costs and the coverage. Figure 2 shows the overlap progress as a function of the number of iterations using different path generation algorithms. From Figure 3 we can see that, a route choice set with 25 routes are sufficient to reach 80% overlap threshold for most of the algorithms, except the simulation methods.

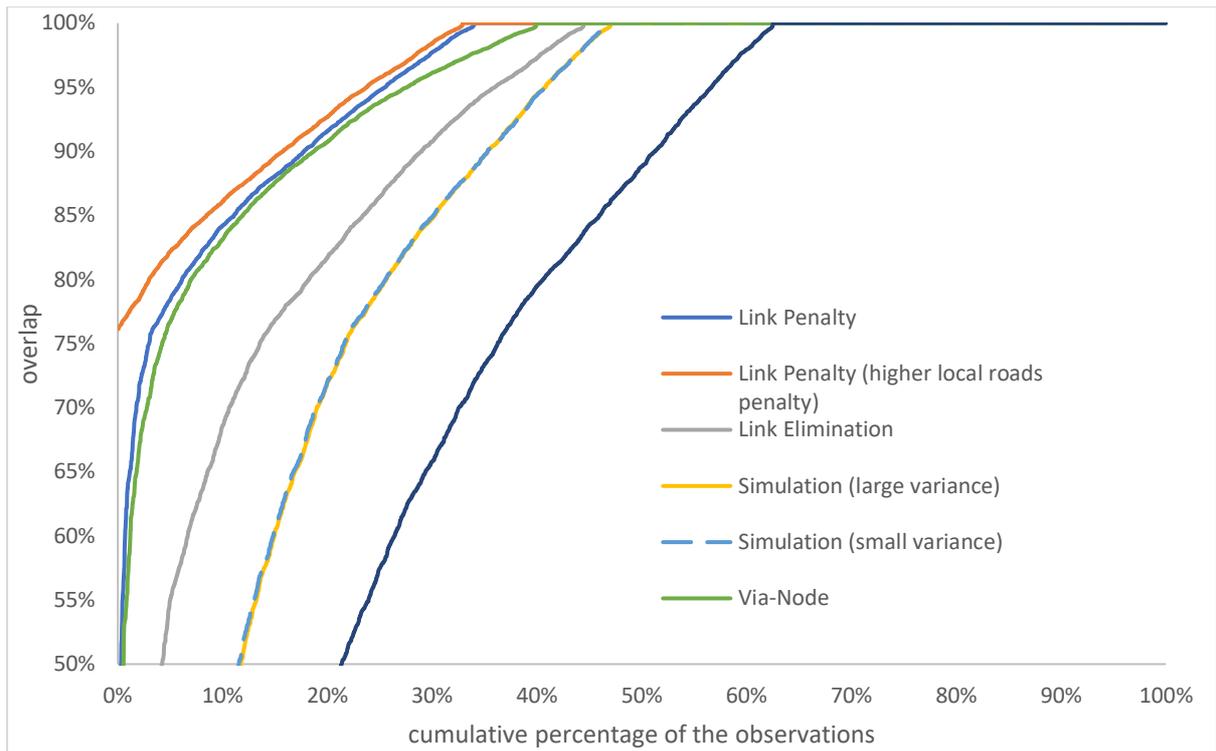

Figure 2 Distribution of Coverage over 6,000 Observations

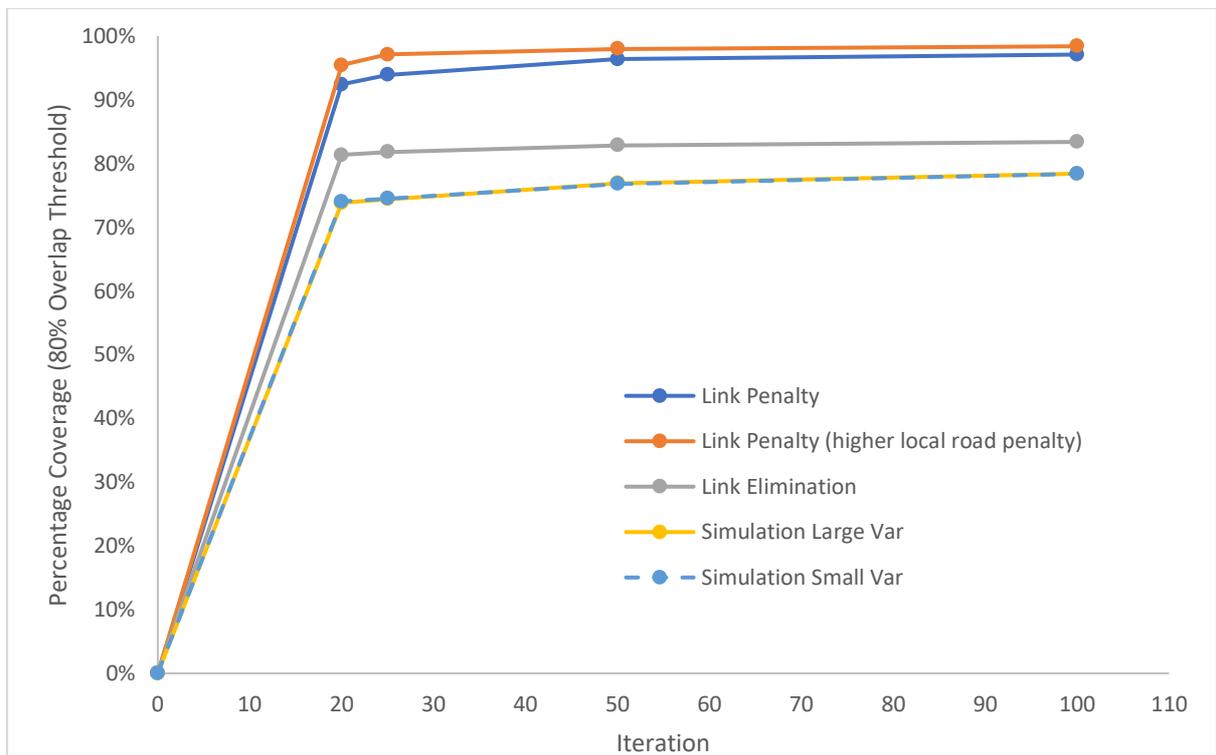

Figure 3 Percentage Coverage with respect to Iterations

We further calculate the ratio between the number of links (different from shortest path) and number of routes (except shortest path), which indicates the heterogeneity of the route choice set. A higher ratio implies that the paths share fewer links and are most likely to be more heterogeneous, Table 4 shows the link/route ratio for different path generation algorithms.

Table 3 Ratio between number of links and number of routes

| | Link Penalty | Link Penalty (higher local road penalty) | Link Elimination | Simulation (small variance) | Simulation (large variance) | Via-Node |
|---|---|---|---|---|---|---|
| # Link / # Route | 7.70 | 7.79 | 8.86 | 12.03 | 12.05 | 8.29 |
| # Link / # Route (exclude shortest path) | 7.24 | 7.40 | 6.87 | 4.09 | 4.32 | 7.67 |

From the table above, we can interpret that via-node method is able to produce routes that are sufficiently different from the shortest path. The route set is more heterogeneous compared to link penalty method with higher local road penalty, because of the overlap constraint presented in via-node method. Note the significant drops regards to simulation methods. Even with a large variance, the simulation generates less unique routes compared to the other methods, and the generated routes are very similar to the shortest path.

# CONCLUSION

This paper evaluates different path generation algorithms using a large GPS trajectory dataset from Tel Aviv metropolitan area. The GPS trajectory set is cleaned and filtered, and map matched to the Tel Aviv network, which results in 6,000 representative observations.

Using only the shortest path labels, almost 60% percent of the 6,000 observations can be covered (assuming a threshold of 80% overlap). This result is significantly higher in comparison to previous results reported in the literature, and suggests compliance with navigation apps.

Two variants of the link penalty methods are studied, in which one of them accounts for preference of using higher hierarchical roads, and outperforms most of the algorithms with 97% coverage (assume 80% overlap threshold) using only 25 iterations.

The via-node method is also evaluated, and the result shows satisfying coverage (94% with 80% overlap threshold). Moreover, the via-node method generates more heterogeneous routes in terms number of links to routes ratio (exclude shortest path) compare to all other methods.

The generated routes with the via-node method will form the basis for route choice model estimation. Further research will compare model estimation results between using an explicit choice set and implicit methods.